\documentclass{mn2e}
\usepackage{graphicx}
\usepackage{epsfig}

\newcommand{\AU}{au}

\newcommand{\gae}
{\lower 2pt \hbox{$\, \buildrel {\scriptstyle >}\over {\scriptstyle \sim}\,$}}
\newcommand{\lae}
{\lower 2pt \hbox{$\, \buildrel {\scriptstyle <}\over {\scriptstyle \sim}\,$}}

\newcommand{\0}{\phantom{0}}

\begin{document}

\title[Saturnian resonances in meteoroid streams]
{Saturnian mean motion resonances in meteoroid streams}

\author[A. Sekhar and D. J. Asher]
{A. Sekhar$^{1,2*}$ and D. J. Asher$^1$\\
 $^1$Armagh Observatory, College Hill, Armagh BT61\ 9DG\\
 $^2$Queen's University of Belfast, University Road, Belfast BT7 1NN\\
 $^*$E-mail: asw@arm.ac.uk , asekhar01@qub.ac.uk \\ }

\date{{\bf Accepted}: 2013 May 16; {\bf Received}: 2013 May 16; {\bf In Original Form}: 2013 Apr 26; {\bf MNRAS Letters}}

\maketitle

\begin{abstract}
Many previous works have shown the relevance and dynamics of Jovian mean
motion resonances (MMR) in various meteoroid streams. These resonant
swarms are known to have produced spectacular meteor displays in the
past. In this work we investigate whether any MMR due to Saturn are
feasible, and subsequently check whether such effects are strong enough
to trap meteoroids so as to cause enhanced meteor phenomena on Earth.
Extensive numerical simulations are done on two major meteoroid streams,
which are known to exhibit exterior Jovian resonances. The roles of the
1:6 and 5:14 Jovian MMR have already been studied in the Orionids and
Leonids respectively. Now we find strong evidence of 1:3 and 8:9
Saturnian MMR in Orionids and Leonids respectively. The presence of
compact dust trails in real space due to these two Saturnian resonances
is confirmed from our calculations.

\end{abstract}


\section{Introduction}

Jovian mean motion resonances (MMR) are known to have played an important
role in determining the long term evolution of many meteoroid streams
(Asher \& Emel'yanenko 2002; Ryabova 2003; Jenniskens 2006; Vaubaillon,
Lamy \& Jorda 2006; Soja et al.\ 2011). In many cases the compact dust trails
due to Jovian resonances have led to observed outbursts and storms
(Astapovich 1968; Yeomans 1981; Asher \& Clube 1993; Rendtel \& Betlem 1993;
Brown et al.\ 2002; Trigo-Rodr\'iguez et al.\ 2007; Rendtel 2008; Kero et
al. 2011). Most of these observational records match with theoretical
simulations to a very good degree. Generally it is seen that the orbital
evolution of resonant meteoroids is dramatically different from that of the
non-resonant ones (Sato \& Watanabe 2007; Christou, Vaubaillon \& Withers
2008). The remarkably different precession rates and particle concentrations
of resonant swarms, in comparison to non-resonant meteoroids, lead to varying
levels of meteor activity in various showers. Hence correlating enhanced
meteor activity with known Jovian resonances (Asher, Bailey \& Emel'yanenko
1999; Jenniskens et al.\ 2007; Rendtel 2007; Sekhar \& Asher 2013) and
subsequent predictions of future meteor outbursts have been a very active
field for some decades. Such calculations have gained very high precision
over the years (Jenniskens et al. 1998; McNaught \& Asher 1999).

Nevertheless we realise that no detailed simulations or analysis were done
regarding resonances in meteoroid streams due to Saturn's gravitational effects
except a brief mention of a possible 8:9 MMR in Leonids (Stoney \& Downing
1898; Brown 2001). Most scientists seem to have presumed that Saturnian
resonances are
either too weak or practically non-existent when it comes to producing 
enhanced meteor phenomena on Earth. Our simulations here show that this
assumption is not true. We find conclusive evidence that strong Saturnian
resonances are feasible as well as effective in trapping large numbers of
meteoroids which can lead to formation of compact dust trails in space. Even
though Saturnian resonances are quite rare compared to Jovian resonances in
the context of known meteoroid streams, the newly
found Saturnian resonances in this work show significant strength and
stability which can in turn relate to spectacular meteor outbursts in the
past and future. This paper investigates such Saturnian resonances in two
major streams which are known to exhibit exterior Jovian resonances
(Emel'yanenko 2001) namely the Orionids and Leonids.

\begin{figure}
(a)\\[-\baselineskip]
\includegraphics[width=\columnwidth]{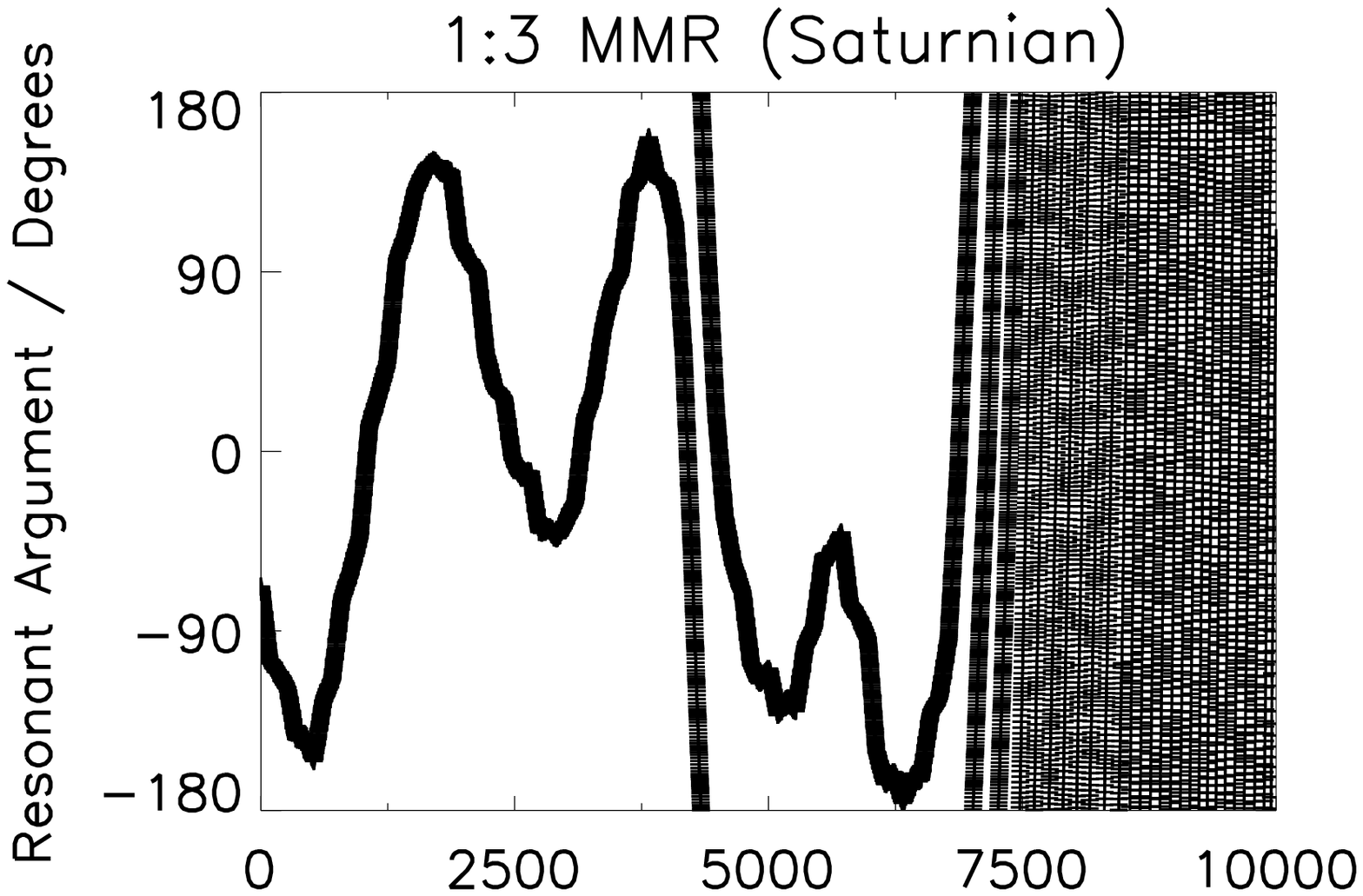}
\\[-5mm]
(b)\\[-\baselineskip]
\includegraphics[width=\columnwidth]{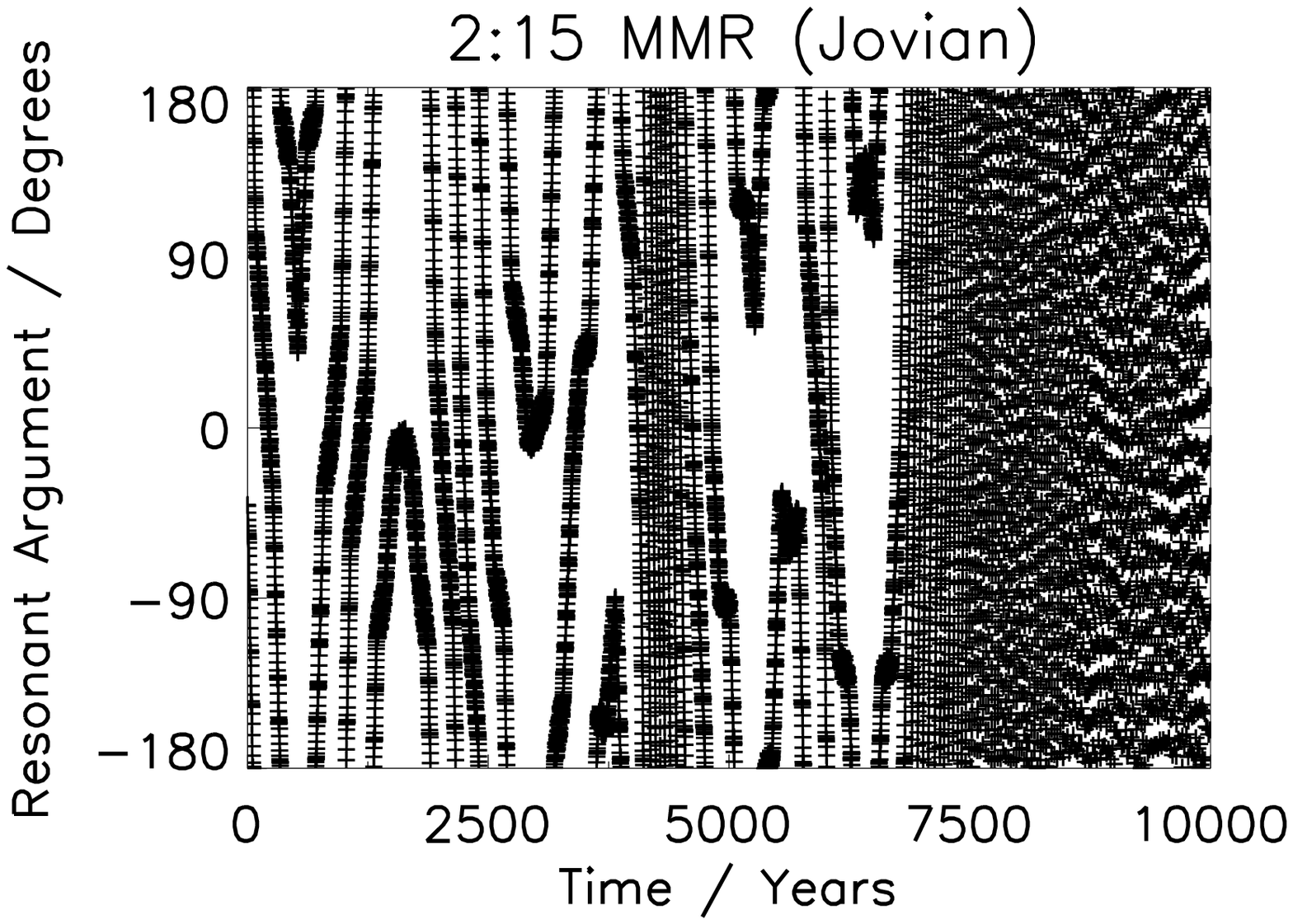}
\caption{(a) Libration of 1:3 (Saturnian) resonant argument for an Orionid
test particle, confirming presence of 1:3 MMR with Saturn.
(b) Circulation of 2:15 (Jovian) resonant argument for the same particle,
confirming absence of 2:15 MMR with Jupiter.}
\label{ORIsigma}
\end{figure}

\begin{table}
\centering
\caption{Order of resonances discussed in this work and applying
combinatorics to calculate resonant arguments permitted by d'Alembert rules. T is the approximate number of successive years for the Earth to encounter a single Leonid/Orionid resonant zone. P is the interval until the next series of successive encounters. In previous work (Sekhar \& Asher 2013) on Orionids, T $\sim$ 5-6 yr \& P $\sim$ 71 yr for 1:6 Jovian, and T $\sim$ 1-2 yr \& P $\sim$ 77 yr for 2:13 Jovian.}
\label{}
\begin{tabular}{@{}rcccc@{}}
\hline
MMR &Order & Number of Possible & P & T  \\
p:(p+q)  &    q         & Resonant Arguments     &(yr) & (yr)\\ 
\hline
8:9 S    & \01        & \0\0\02  &33   & \02   \\
1:3 S    & \02        & \0\0\06  &88  &   22 \\
5:14 J   & \09        &   \0110  &33   &  \01 \\
2:15 J   &  13        &   \0280   &-  &   -\\
16:45 J  &  29        &    2480  &-    &  -\\
\hline
\end{tabular}\\
\end{table}

\section{Separating Jovian and Saturnian Resonances}
\label{sigma}

Since there is a well known 2:5 near commensurability, widely known as the
great inequality (Kepler 1672; Halley 1676; Laplace 1785; Hill 1890; Lovett 1895;
Brouwer \& van Woerkom 1950; Milani \& Kne\v{z}evi\'c 1990), between the
orbits of Jupiter and Saturn, it is vital to cross check the evolution of
resonant arguments (for any particular resonance ratio involving Saturn) of
the meteoroid particle for these two nearby resonances simultaneously. For
example in the case of 1:3 MMR (Saturnian), multiplying this ratio by 2:5
gives the nearby 2:15 MMR (Jovian). The current investigation confirms the
presence of 1:3 Saturnian MMR (semi-major axis $a_{n}=19.84$ \AU) in the
Orionids. It is essential to check the adjacent 2:15 Jovian resonant
argument ($a_{n}= 19.93$ \AU) to avoid any misleading signature from this
nearby Jovian MMR. Here we use $a$ = $a_{n}$ = the `nominal resonance
location' (Murray \& Dermott 1999, section 8.4) for exterior resonances
(Peale 1976) of the form p:(p+q) where q is the order of resonance and
repeated conjunction occurs for every p orbits of the particle.

Figure \ref{ORIsigma}(a) shows the 1:3 Saturnian resonant argument $\sigma$
librating continuously for about 4 kyr, then briefly becoming non-resonant
(overall range of $\sigma$ becomes 360\degr\ for a short while) and
subsequently falling into the same resonance for a further 2 kyr, in the case
of an Orionid test particle. Figure \ref{ORIsigma}(b) shows the 2:15
resonant argument (Jovian) clearly circulating during the same time frame.
The starting epoch is JD 1208900.18109 = 1404 B.C. October 15.68109, the
oldest credible computed perihelion passage time of 1P/Halley (Yeomans \&
Kiang 1981).

\begin{figure}
(a)\\[-\baselineskip]
\includegraphics[width=\columnwidth]{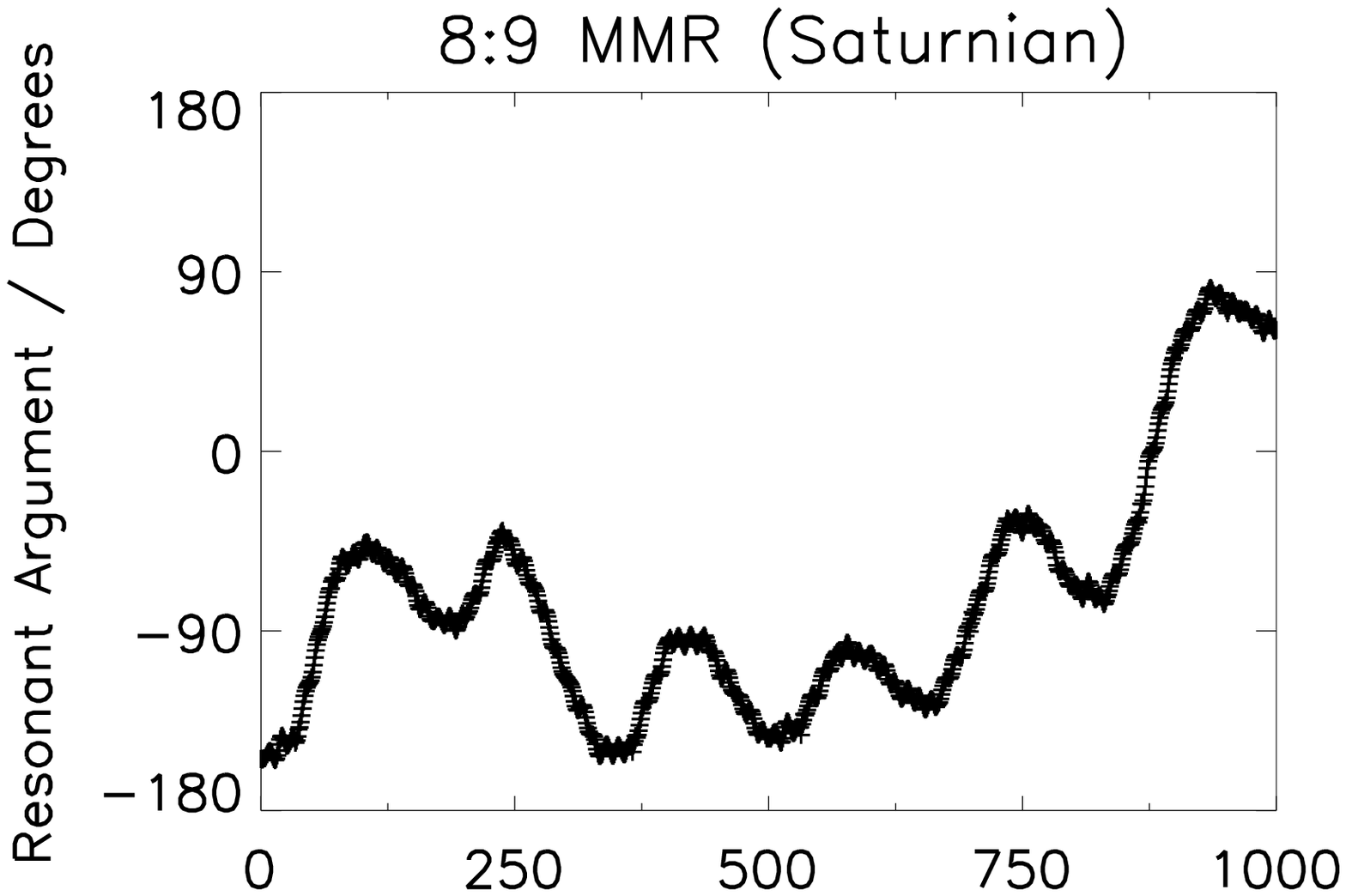}
\\[-5mm]
(b)\\[-\baselineskip]
\includegraphics[width=\columnwidth]{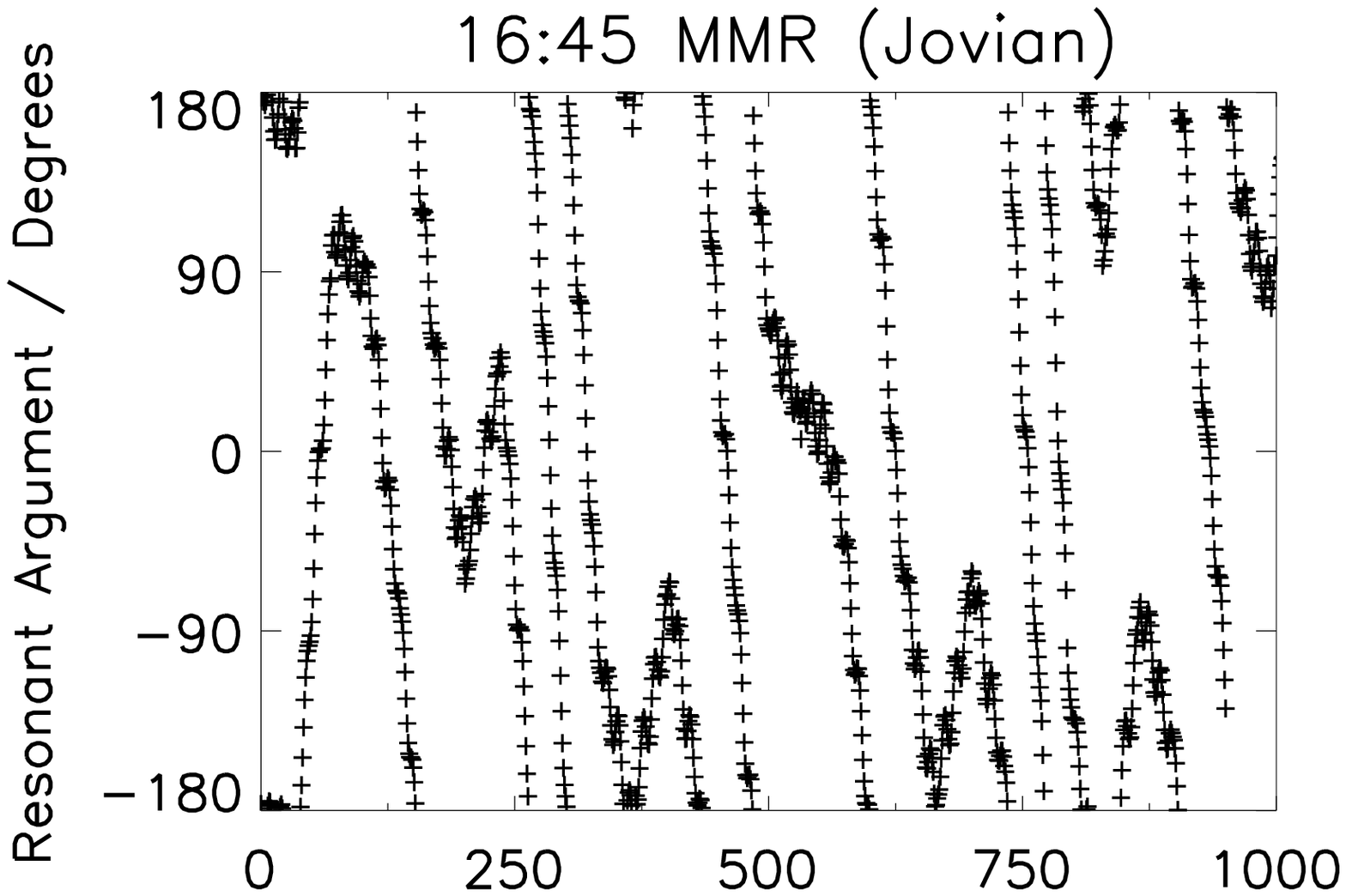}
\\[-5mm]
(c)\\[-\baselineskip]
\includegraphics[width=\columnwidth]{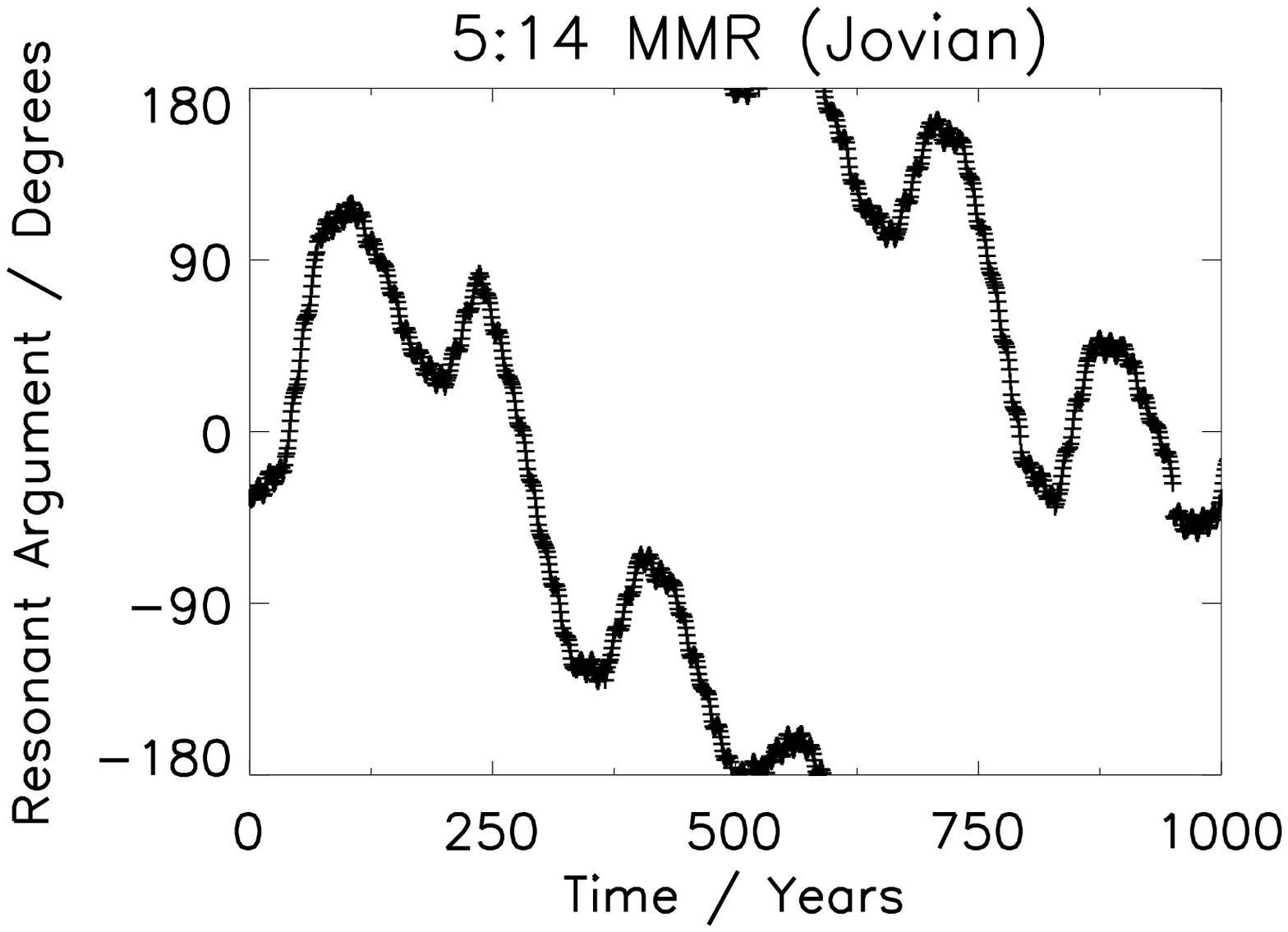}
\caption{(a) Libration of 8:9 (Saturnian) resonant argument for a Leonid test
particle, confirming presence of 8:9 MMR with Saturn.
Circulation of (b) 16:45 and (c) 5:14 Jovian resonant arguments for the same
particle confirm absence of 16:45 and 5:14 MMR with Jupiter.}
\label{LEOsigma}
\end{figure}

For the Leonids we confirm the presence of 8:9 Saturnian MMR ($a_{n}=10.32$
\AU). Figure \ref{LEOsigma}(a) shows $\sigma$ for this MMR librating for
$\sim$700 yr before starting to drift from the libration centre (initial
epoch = JD 2220280.1685 = computed return time of
55P/Tempel-Tuttle in 1366). Figures \ref{LEOsigma}(b) and (c) show the
adjacent 16:45 ($a_{n}=10.36$ \AU) and 5:14 ($a_{n}=10.33$ \AU) resonant
arguments (Jovian) clearly circulating for the same test particle during the
same time frame. Multiplying 8:9 MMR (Saturnian) by 2:5 gives the nearby
16:45 MMR (Jovian). In terms of $a_{n}$, 5:14 MMR (Jovian) is even nearer
than 16:45 is to 8:9 MMR (Saturnian). Hence both these Jovian cases were
verified to get a comprehensive conclusion.

In order to absolutely confirm the presence of Saturnian MMR and rule out the
presence of nearby Jovian MMR, many of the different possible combinations
(see Table 1) of 1:3, 8:9 and 2:15, 16:45, 5:14 resonant arguments allowed by
d'Alembert rules (Murray \& Dermott 1999, sections 6.7 and 8.2) were checked
to confirm libration and circulation respectively (cf.\ Sekhar \& Asher
2013). The resonant arguments plotted in Figures \ref{ORIsigma}(a)
and \ref{LEOsigma}(a) are respectively
$\sigma = \lambda_s-3\lambda_m+2\varpi_m$ and
$\sigma = 8\lambda_s-9\lambda_m+\varpi_m$, where $\lambda$ and $\varpi$
denote mean longitude and longitude of pericentre, subscripts $s$ and $m$
standing for Saturn and meteoroid particle. Because both Halley's and Tempel-Tuttle's orbits are retrograde, here we use the modified definition of $\varpi = \Omega - \omega$ (Saha \& Tremaine 1993; Whipple \& Shelus 1993).

These techniques clearly show that Saturnian resonances are indeed real and
not entwined with the near commensurate Jovian resonances. In
Section \ref{geom} we integrate ranges of particles to show dense clusters of
Saturnian resonant meteoroids in space retaining their compact structure for
many kyr.

The numerical integrations in this work were done using the \textsc{mercury}
package (Chambers 1999) implementing the \textsc{radau} algorithm (Everhart
1985) with accuracy parameter set to $10^{-12}$ and including the sun and
eight planets, whose orbital elements were retrieved from JPL Horizons
(Giorgini et al.\ 1996). Elements for the parent bodies 1P/Halley and
55P/Tempel-Tuttle were taken from Marsden \& Williams (2008). Radiation
pressure and Poynting-Robertson effects were not included in any
integrations.

\section{Geometry of Resonant Zones and Ecliptic Plane Crossings}
\label{geom}

\subsection{Orionids}
\label{ORIgeom}

\begin{figure}
(a)\\[-\baselineskip]
\includegraphics[width=\columnwidth]{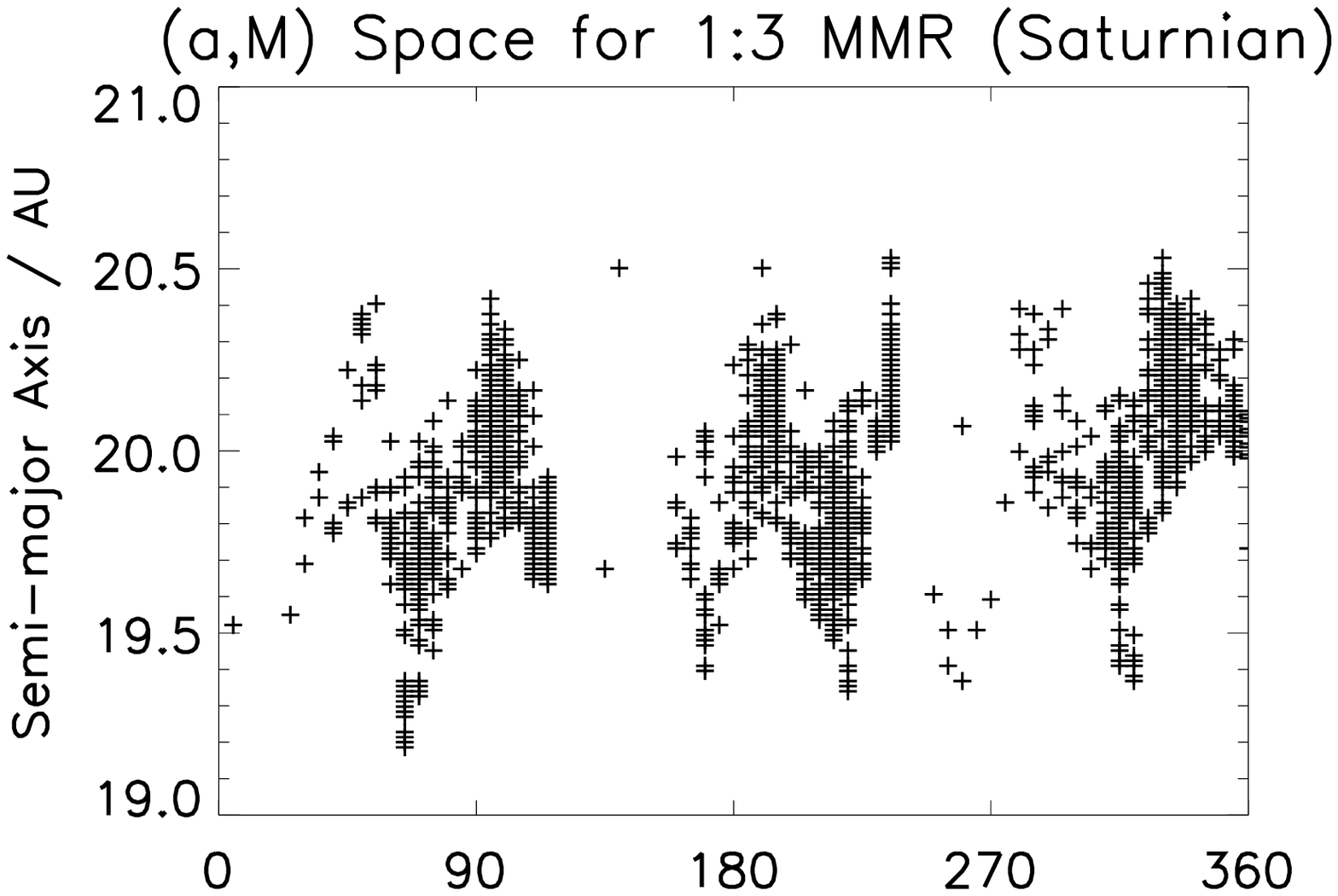}
\\[-5mm]
(b)\\[-\baselineskip]
\includegraphics[width=\columnwidth]{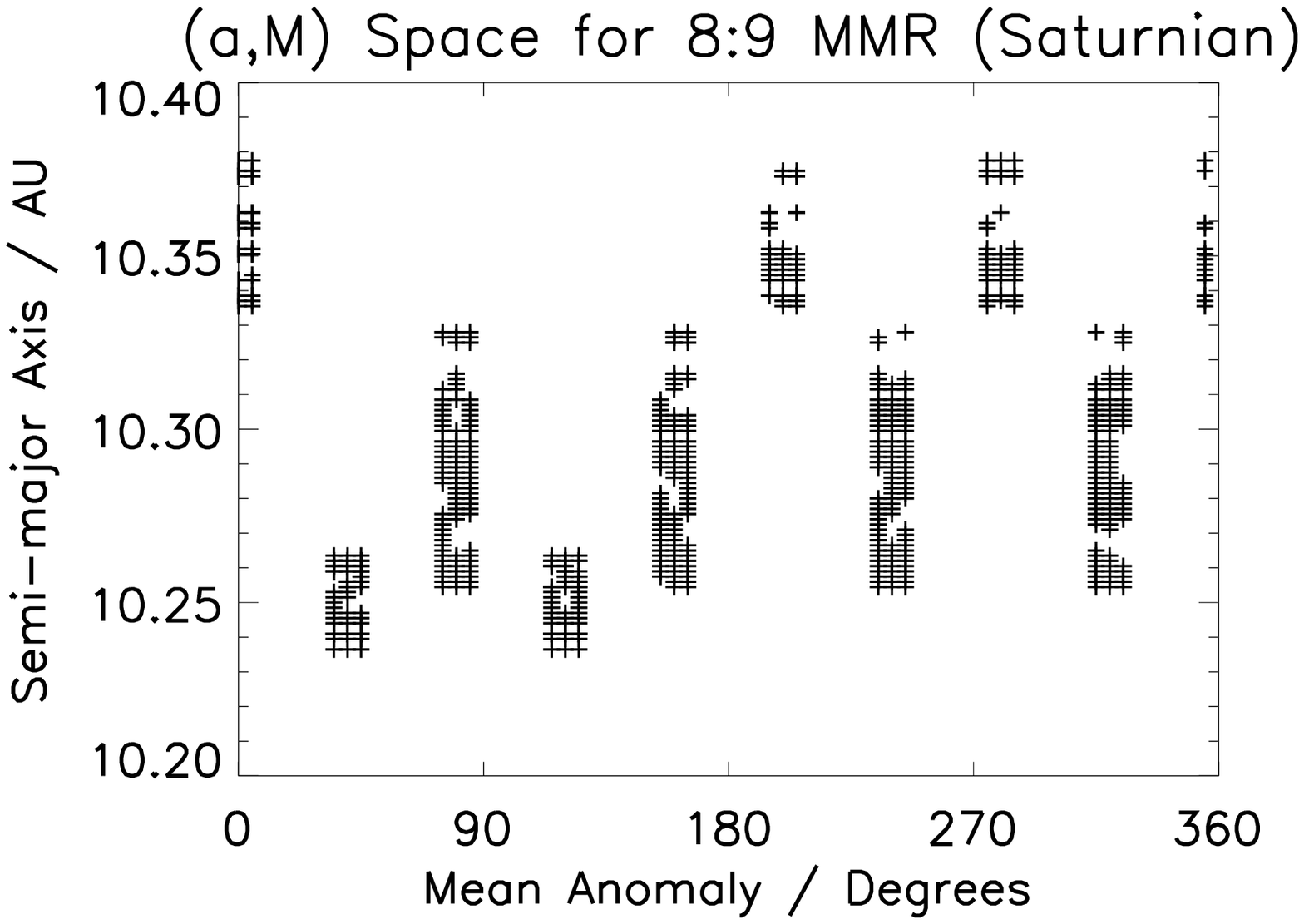}
\caption{(a) three resonant zones for 1:3 Saturnian MMR in Orionids as
a function of $a$ and $M$ at initial epoch JD 1208900.18109;
(b) nine zones for 8:9 Saturnian MMR in Leonids as a function of ($a$,$M$) at
initial epoch JD 2220280.1685.}
\label{aM}
\end{figure}

Figure \ref{aM}(a) shows the general picture of resonant zones for the 1:3
Saturnian MMR ($a_{n}=19.84$ \AU): 7200 particles were integrated forward
from 1404 B.C., varying the initial $a$ from 19.0 to 20.8 \AU\ in steps of
0.018 \AU, and initial $M$ from 0 to 360\degr\ in steps of 5\degr, keeping
other elements (namely $q$, $i$, $\omega$ and
$\Omega$) the same as 1P/Halley. Particles are plotted that librate
continuously for 4 kyr. The distribution shows their initial semi-major axis
$a$ and mean anomaly $M$. Resonant particles were identified (cf.\
Section \ref{sigma}) by a simple algorithm which looks at the overall range
of resonant argument (for different combinations allowed by d'Alembert rules)
for each particle every 10 yr during the whole 4 kyr. A snap shot of the same
($a$,$M$) phase space 4 kyr later shows a similar picture with three dense
clouds of resonant particles. Our simulations show that 1:3 resonant
meteoroids can retain compact structures for many kyr (4 kyr is typical). A
resonant meteor outburst is a possibility if Earth passes through one of
these three clumps in space. When the Earth misses these dense clouds, a
normal meteor shower can still occur.

The $a$ range spanned by the resonant zone (Fig.\ \ref{aM}a) is equivalent to
perihelion tangential ejection speeds in the range of about +40 to +65
m\,s$^{-1}$ at the 1404 B.C. return of 1P/Halley. These ejection velocities
are realistic in cometary activity (Whipple 1951). Moreover radiation
pressure acts in the same way as positive (= forward) ejection velocities,
i.e.\ increasing the orbital period, and for visual meteor sized meteoroids
the effect of radiation pressure on the period is quantitatively comparable
to these ejection speeds (cf.\ Kondrat'eva \& Reznikov 1985; Williams 1997;
Asher \& Emel'yanenko 2002). Hence these positive ejection velocities (in the
gravitational integration model) required to populate this 1:3 resonant zone
at this epoch imply that significant numbers of real meteoroids were released
by 1P/Halley into this resonance. Over centuries the comet will drift through
the three 1:3 resonant zones and populate all of them.

\begin{figure}
(a)\\[-\baselineskip]
\includegraphics[width=\columnwidth]{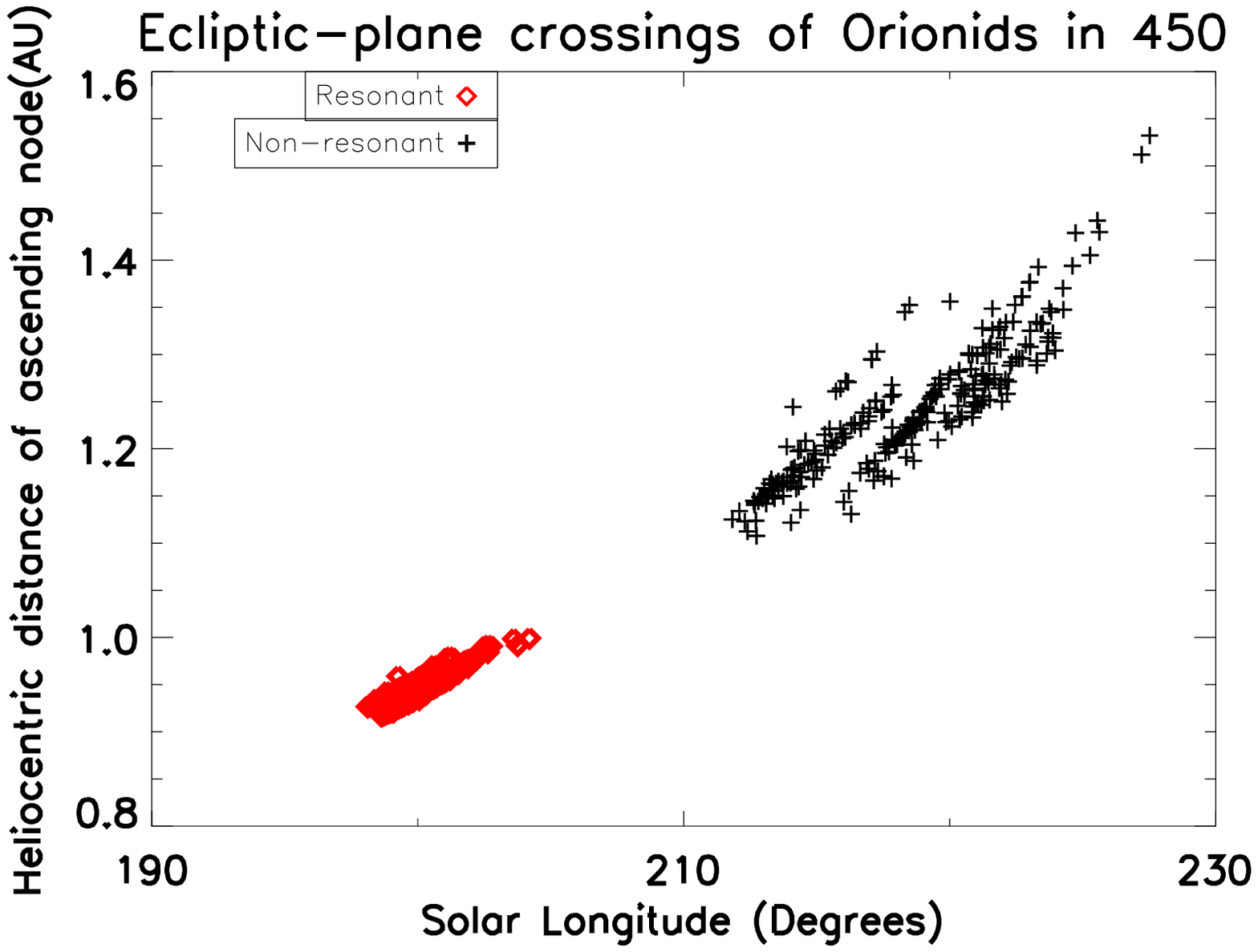}
\\[-1mm]
(b)\\[-\baselineskip]
\includegraphics[width=\columnwidth]{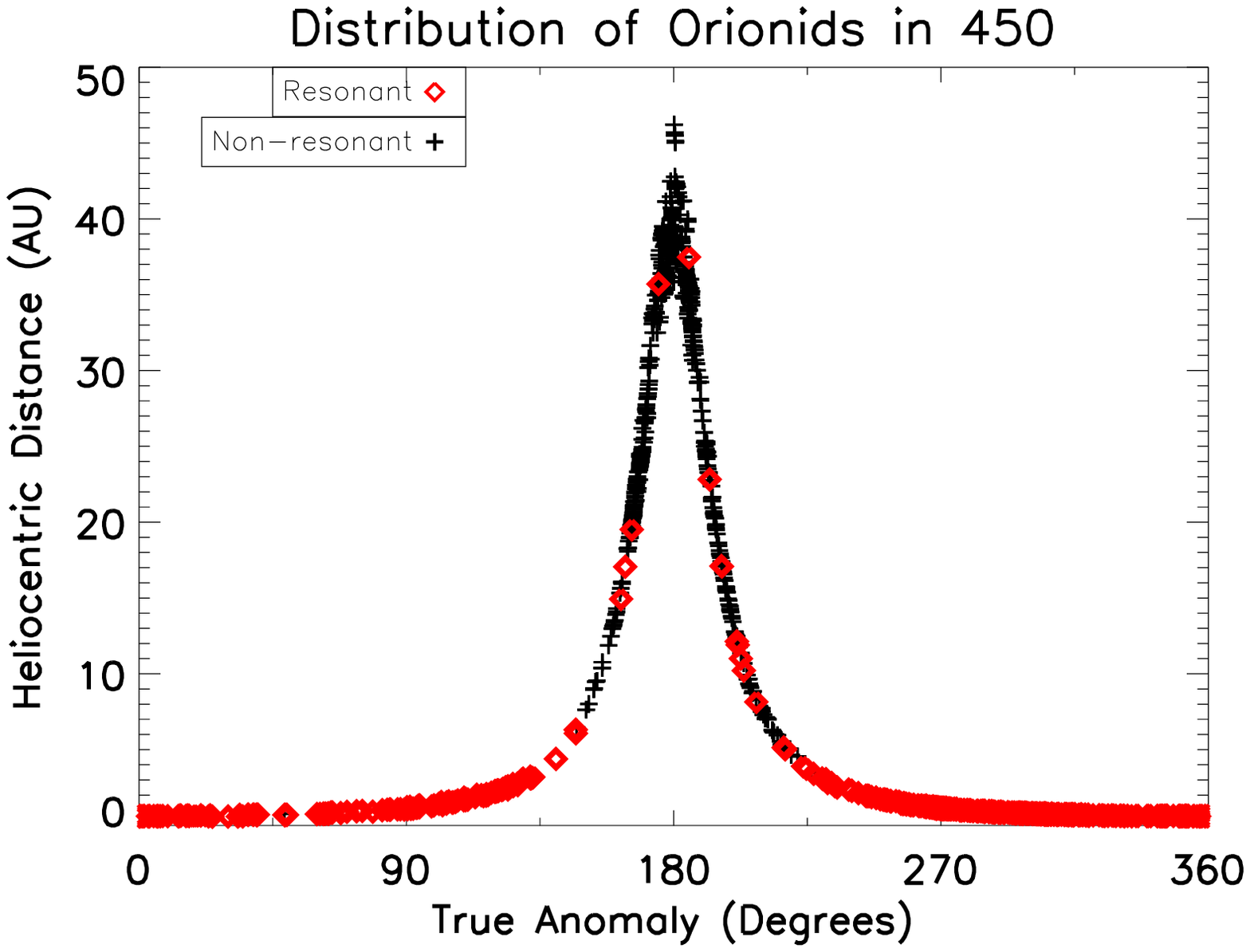}
\caption{(a) Ecliptic plane crossings; and (b) distribution of
heliocentric distances, for sets of 1:3 Saturnian MMR and non-resonant
Orionid particles in 450 A.D. Both sets of particles evolved
dynamically for 2 kyr but the dense clustering of resonant meteoroids
contrasts with the large dispersion of non-resonant particles.
}
\label{ORIecl}
\end{figure}

\begin{figure}
(a)\\[-\baselineskip]
\includegraphics[width=\columnwidth]{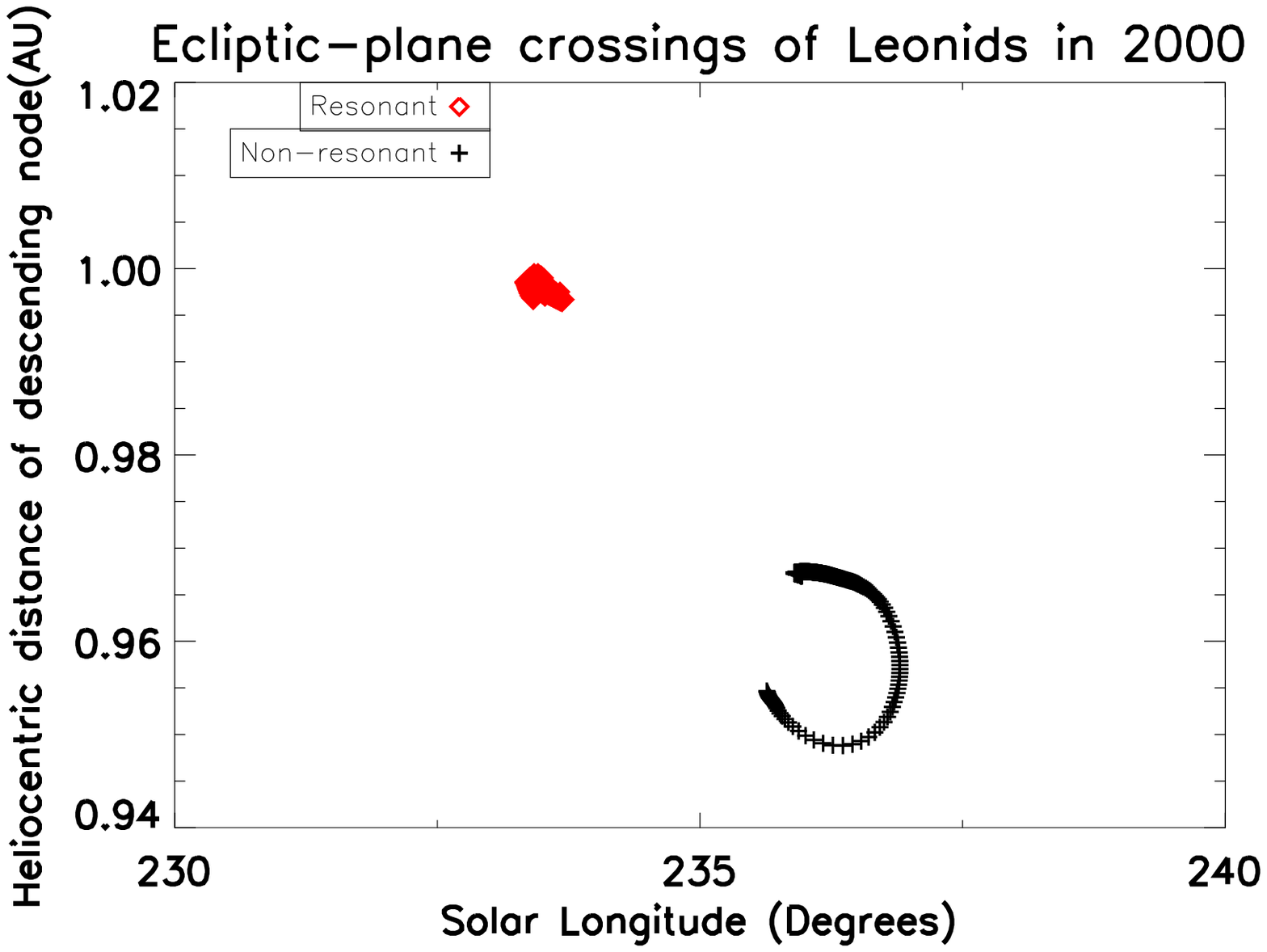}
\\[-1mm]
(b)\\[-\baselineskip]
\includegraphics[width=\columnwidth]{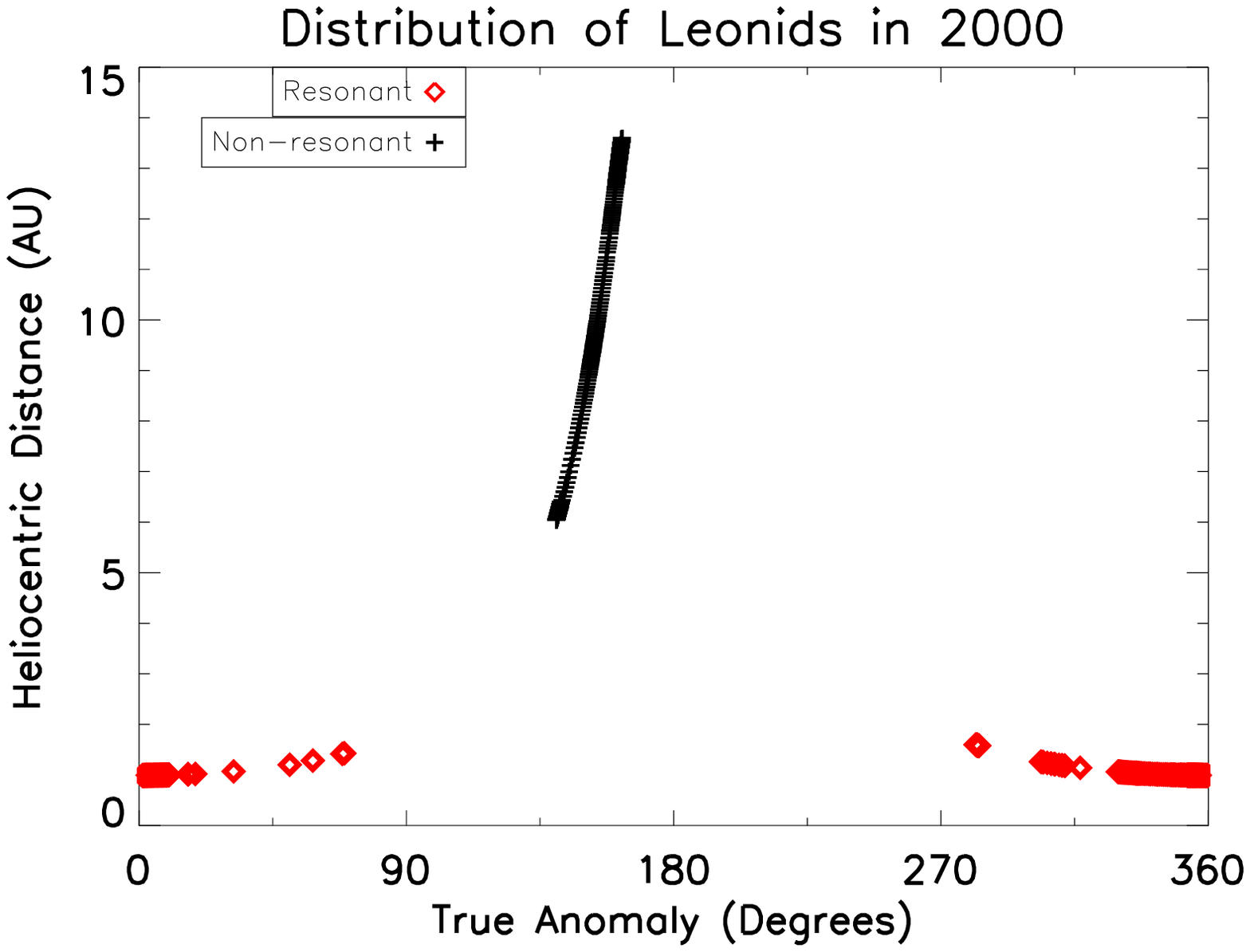}
\caption{(a) Ecliptic plane crossings; and (b) distribution of
heliocentric distances, for sets of 8:9 MMR (Saturnian) and non-resonant
Leonid particles in 2000 A.D., both sets having evolved over the same
time frame. The resonant case shows very compact dust trails whereas
significant dispersion is indicated in the non-resonant case.}
\label{LEOecl}
\end{figure}

The active role of 1:6 and 2:13 Jovian MMR in causing strong Orionid meteor
outbursts has previously been demonstrated (Rendtel 2007; Sato \& Watanabe
2007; Sekhar \& Asher 2013). The 1:3 Saturnian MMR can produce similarly
compact structures in the Orionid stream. Figure \ref{ORIecl}(a) shows
the (ascending) nodal crossing distribution for particles from a single
1:3 resonant zone, contrasted with non-resonant particles (cf.\
librating and circulating model comparisons of Emel'yanenko \& Bailey
1996, e.g.\ their fig.\ 3). In the resonant case, 2000 particles were
integrated from Halley's 1404 B.C. return ($M$=0), varying the initial $a$ from
20.0 to 20.1 in steps of $ 5 \times 10^{-5}$ \AU\ and keeping $q$, $i$, $\omega$
and $\Omega$ the same as the comet.
All parameters were the same for the non-resonant particles except the
starting epoch was adjusted by $\sim$2 yr ($M$=0 at JD 1208171.10151 was
chosen) so that the evolution of non-resonant Orionids can be studied
for the same time frame. There were no significant close encounters with
planets during this extra integration time. The very compact structures
in Fig.\ \ref{ORIecl} for the librating particles prove the physical
presence of dense meteoroid concentrations in space. Moreover the
particle distribution along the orbit shows the non-resonant ones to be
dispersed over a very large range of heliocentric distances which in
turn leads to very low meteoroid concentrations.

\subsection{Leonids}

To explore the 8:9 MMR ($a_{n}=10.32$ \AU) in the Leonid stream, we
integrated 7200 particles varying initial $a$ from 10.16 to 10.46 \AU\
in steps of 0.0030 \AU\ and $M$ in steps of 5\degr. Figure \ref{aM}(b)
shows the distribution of particles with various values of initial
($a$,$M$) that have not circulated through 360\degr\ for the entire
period of 1000 years starting from 1366 A.D. Virtually all such
particles are resonant for at least 700 yr (Fig.\ \ref{LEOsigma}a, and
similar plots for other particles not shown).
The inherent mechanism of resonant zones is the same as described for
Orionids (Section \ref{ORIgeom}) except that there are nine dense clumps
in this case. A snap shot of the same phase space after 700 years shows
a similar picture with nine clumps of resonant particles. Our
simulations show that 8:9 resonant meteoroids can retain compact
structures for up to many centuries (typically $\sim$700 yr; cf.\
Fig.\ \ref{LEOsigma}a).

Meteoroid ejection velocities in the range of about $-$20 to $-$4
m\,s$^{-1}$, perfectly feasible during outgassing activity, can populate
the entire resonant zone (shown in Fig.\ \ref{aM}b) at the 1366 return
of 55P/Tempel-Tuttle. The requirement for negative ejection velocities
(in a gravitational integration model) will have an opposite effect to
that discussed in Section \ref{ORIgeom}, i.e.\ the real population of
resonant meteoroids will consist of particles less affected by radiation
pressure. This would mean enhanced chances of narrow trails of larger
meteoroids in turn leading to brighter meteors when they intersect
Earth.

Asher et al.\ (1999) and Brown \& Arlt (2000) have shown the relevance
of 5:14 Jovian MMR in causing intense meteor outbursts in the recent
past. Figure \ref{LEOecl}(a) shows the (descending) nodal crossing
distribution in 2000 A.D. for meteoroids from a single 8:9 (Saturnian)
resonant zone and for non-resonant meteoroids. The resonance leading to
the compact distribution is similar to the mechanism discussed in
Section \ref{ORIgeom}. Also as in Section \ref{ORIgeom} the non-resonant
meteoroids are dispersed over a much larger range of heliocentric
distance (Fig.\ \ref{LEOecl}b). Hence these stable resonances can play
an important role when it comes to spectacular meteor outbursts.

For Fig. \ref{LEOecl}, the integrations were of 2000 particles with
initial $a$ from 10.35 to 10.36 in steps of $5 \times 10^{-6}$ \AU\ and
$q$, $i$, $\omega$ and $\Omega$ the same as Tempel-Tuttle. All the
particles were integrated for 700 years from the 1366 A.D. return ($M$=0).
The non-resonant set had the same initial conditions except that, as in
Section \ref{ORIgeom}, the starting epoch was offset so that the
dynamics of non-resonant Leonids can be analysed ($M$=0 at JD
2219597.24930 was chosen). No relevant close encounters occurred during
this additional integration time.

\vspace*{-5mm}

\section{Conclusions}

We have shown that Orionid meteoroids can stay continuously in 1:3 MMR with
Saturn for $\sim$4 kyr and Leonid meteoroids in 8:9 MMR with Saturn for
$\sim$700 yr. It is verified that none of these resonant signatures are due to
nearby Jovian resonances such as 2:15 and 5:14 in Orionids and Leonids
respectively. The survival times (of the order of 10$^3$ yr) and density
distributions of these Saturnian resonances, which can in turn lead to very
compact dust trails producing enhanced meteor activity, are comparable to
those due to previously known Jovian resonances like 2:13 MMR in the Orionids
discussed in Sekhar \& Asher (2013).

Generally the lower the order q of a resonance, the higher its strength. In
the resonances mentioned above, it is then fair to assume that the weaker
gravitational effect (in comparison to Jupiter) of Saturn is compensated by
the difference in order of resonance. In 8:9 and 1:3 MMR (Saturnian) q is 1
and 2 respectively, very low compared to 9 and 13 in the case of 5:14 and
2:15 MMR (Jovian) respectively. Saturn's effects can become very relevant
and significant in such cases. Hence one cannot rule out the possibility of
an interesting time frame (in past or future) when there could be spectacular
Orionid or Leonid meteor displays due to Saturn's effects comparable to those
due to Jovian effects.

\vspace*{-5mm}

\section*{Acknowledgments}

The authors thank J\"urgen Rendtel for helpful comments and express their gratitude to the Department of Culture, Arts and Leisure of
Northern Ireland for the generous funding to pursue astronomical research at
Armagh Observatory.

\end{document}